\documentclass[12pt]{article}
\usepackage{psfig}

\def\lesssim{\mathrel{\mathpalette\vereq<}}
\def\gtrsim{\mathrel{\mathpalette\vereq>}}
\makeatletter
\def\vereq#1#2{\lower3pt\vbox{\baselineskip1.5pt \lineskip1.5pt
\ialign{$\m@th#1\hfill##\hfil$\crcr#2\crcr\sim\crcr}}}
\makeatother
\begin{document}

\begin{titlepage}
\begin{center}
\today    \hfill    LBNL-44351 \\
~{} \hfill UCB-PTH-99/46  \\
~{} \hfill hep-ph/9910286\\

\vskip .1in

{\large \bf Earth Matter Effect in\\$^7$Be Solar Neutrino
  Experiments}\footnote{
  This work was supported in part by the Director, Office of
  Science, Office of High Energy and Nuclear Physics, Division of High
  Energy Physics of the U.S. Department of Energy under Contract  
  DE-AC03-76SF00098 and in part by the National Science Foundation
  under grant PHY-95-14797.  HM was also supported by the Alfred
  P. Sloan Foundation and AdG by CNPq (Brazil).}

\vskip 0.3in

Andr\'e de Gouv\^ea,\footnote{New address from Oct 1, 1999: Theory
  Division, CERN, CH 1211, Geneva, Switzerland.} Alexander Friedland,
  and Hitoshi Murayama

\vskip 0.05in

{\em Theoretical Physics Group\\
     Ernest Orlando Lawrence Berkeley National Laboratory\\
     University of California, Berkeley, California 94720}

\vskip 0.05in

and

\vskip 0.05in

{\em Department of Physics\\
     University of California, Berkeley, California 94720}

\end{center}

\vskip .1in

\begin{abstract}
We determine the sensitivity of the KamLAND and Borexino experiments
to the neutrino regeneration effect in the Earth as a function of
$\Delta m^2$ and $\theta$, using realistic numbers for the signal and
background rates. We compare the results obtained with the $\chi^2$
method with those obtained from the conventional day-night asymmetry
analysis.  We also investigate how well one should be able to measure
the neutrino oscillation parameters if a large day-night asymmetry is
observed, taking the LOW solution as an example.  We present an
enlarged parameter space, which contains mixing angles greater than
$\pi/4$ where the {\it heavy} mass eigenstate is predominantly
$\nu_e$, and determine the electron neutrino survival probability for
this traditionally neglected scenario. We emphasize that this portion
of the parameter space yields different physics results when dealing
with the MSW solutions to the solar neutrino puzzle and should not be
neglected.
\end{abstract}

\end{titlepage}

\newpage
\setcounter{footnote}{0}
\section{Introduction}

A number of experiments \cite{Kamiokande,GALLEX,SAGE,Cl,Super-K} have
accumulated over the years a large amount of solar neutrino data. The
data indicate that the number of solar neutrino induced events is
significantly smaller than expected and, furthermore, that the
electron neutrino survival probability is energy dependent. This
``solar neutrino puzzle'' is best solved by assuming that the
electron-type neutrino ($\nu_e$) oscillates into another active
neutrino species (some linear combination of the muon-type neutrino
($\nu_{\mu}$) and the tau-type neutrino ($\nu_{\tau}$)), or a sterile
(weak isosinglet) neutrino. In light of the very robust
Super-Kamiokande evidence for $\nu_{\mu}$ atmospheric neutrino
oscillations \cite{atmospheric}, the oscillation of solar neutrinos
seems a very likely and natural hypothesis.

The current experimental situation is such that there are four
disconnected regions in the two-neutrino oscillation parameter space
that fit the data. One of them, the ``just-so'' solution, relies on
vacuum neutrino oscillations with a very long wavelength (comparable
to the Earth-Sun distance) \cite{bksreview}, while the other three
\cite{bksreview,rate_analysis} rely on the MSW effect \cite{MSW} to
produce the required energy dependence of the electron neutrino
survival probability. Discriminating among all these solutions is the
goal of the current and the next generations of neutrino experiments.

Even though one can classify the solar neutrino puzzle as strong
evidence for neutrino oscillations, it is as yet not considered
definitive. The main foci of criticism traditionally have been that
the Standard Solar Model (SSM) \cite{SSM} might not be accurate enough
to precisely predict the fluxes of different energy components of
solar neutrinos, and that the evidence for solar neutrino oscillations
relies on a combination of hard, different experiments.  Even though
it seems very unlikely that reasonable modifications to the SSM alone
can explain the current solar neutrino data (see, for example,
\cite{ssm_indep}), one still cannot completely discount the
possibility that a combination of unknown systematic errors in some of
the experiments and certain modifications to the SSM could conspire to
yield the observed data. To conclusively demonstrate that there is
indeed new physics in solar neutrinos, the experiments now are aiming
at detecting ``smoking gun'' signatures of neutrino oscillations,
such as an anomalous seasonal variation in the observed neutrino flux or a
day-night variation due to the regeneration of electron neutrinos in the
Earth. In this paper we study the sensitivity reach of two upcoming
neutrino experiments, Borexino and KamLAND, to the Earth regeneration
effect.

Out of all solar neutrino components, both experiments will be most
sensitive to $^7$Be neutrinos. These are neutrinos produced in the
electron capture by $^7$Be nuclei in the Sun's core
($^7$Be$+e^{-}\rightarrow ^7$Li $+\nu_e$). One very important fact is
that these neutrinos are almost monochromatic, with
$E_{\nu}=0.862$~MeV (90\% of the time) or $E_{\nu}=0.383$~MeV (10\% of
the time), depending on the final state of the $^7$Li nucleus. Since
the $E_{\nu}=0.383$~MeV neutrinos cannot be cleanly seen in future
detectors, we will only consider the $E_{\nu}=0.862$~MeV neutrinos,
which will be referred to as the $^7$Be neutrinos.

The study of the $^7$Be neutrino flux is particularly important, for a
variety of reasons. First, in the SSM independent analysis of the
solar neutrino data \cite{ssm_indep}, where one arbitrarily rescales
the flux of neutrinos from different sources, the flux of $^7$Be
neutrinos comes out extremely suppressed (in fact the best fit value
for the $^7$Be flux is negative!), and the measurement of a reasonable
flux would dramatically constrain such attempts. Second, since the
prediction of one particular MSW solution (the small angle solution)
for the survival probability of $^7$Be neutrinos is very different
from the other two solutions, one can separate if from the other two
by measuring the $^7$Be solar neutrino flux. Third, as was recently
shown \cite{seasonal}, one can either establish or exclude the
``just-so'' solution by analyzing the seasonal variation of the $^7$Be
solar neutrino flux at Borexino or KamLAND. Finally, it might also be
possible to separate the $\nu_e$ from the $\nu_{\mu,\tau}$ component
in the $^7$Be flux, by studying the kinetic energy spectrum of recoil
electrons \cite{recoil_spectrum} in future experiments.

It has been known for over a decade that the propagation of solar
neutrinos through the Earth can result in a measurable variation in
the observed neutrino event rates \cite{day_night_all}. Reference
\cite{day_night}, in particular, contains a detailed analysis of the
expected day-night asymmetry for the Super-Kamiokande, Borexino, and
SNO experiments. In this paper we extend the previous analyses in
several important aspects. First, we present an {\it enlarged parameter
space}, where the vacuum mixing angle is allowed to vary over its
entire physical range from 0 to $\pi/2$\footnote{This enlarged parameter
 space has already been mentioned in the context of three-flavor oscillations 
\cite{FLM}.}. 
We find that not only does
the day-night asymmetry stay nonvanishing at maximal mixing
($\theta=\pi/4$), in 
agreement with \cite{Guth_Randall}, but that it also smoothly extends into
the other part of the parameter space ($\pi/4<\theta\leq\pi/2$). Second, we
display the sensitivity regions of KamLAND and Borexino in
this enlarged parameter space, using realistic numbers for the signal
and background rates. In our analysis we use the $\chi^2$ method, and
study the effect of various binning schemes. Finally, we explore the
possibility of using the neutrino regeneration data at the two
experiments in question to measure the oscillation parameters.

This paper is organized as follows.  In Sec.~\ref{dn_effect} we review
the day-night effect and present the day-night asymmetry expected for
$^7$Be neutrinos as a function of the two neutrino oscillation
parameter space. We also introduce an enlarged parameter space,
$0\leq\theta\leq\pi/2$. In Sec.~\ref{KamLAND}, we study the
sensitivity of the KamLAND and Borexino experiments to the day-night
asymmetry and to the zenith angle dependence of the $^7$Be flux. In
Sec.~\ref{measurement} we study the possibility of measuring the
oscillation parameters if a significant day-night effect is observed
at either Borexino or KamLAND. We contrast the analysis of the
day-night asymmetry with the zenith angle distribution. In
Sec.~\ref{conclusion} we present a summary of our results and
conclusions.

\setcounter{equation}{0}
\section{Electron Neutrino Regeneration in the Earth} 
\label{dn_effect}

As was realized over a decade ago \cite{MSW}, neutrino-matter
interactions can dramatically affect the pattern of neutrino
oscillations. The reason for this is that neutrino-matter
interactions are flavor dependent, given that the matter distributions
of interest (the Earth, the Sun) contain only first generation
particles. One well-known consequence of this is that, in the case of
neutrinos produced in the Sun's core, it is possible to obtain an almost
complete $\nu_e\rightarrow \nu_{\rm other}$ transformation even when
the vacuum mixing angle is very small \cite{MSW}. 

It has also been pointed out by several authors \cite{day_night_all,
day_night, Guth_Randall} that matter effects might also be relevant
for neutrinos traversing the Earth. One experimental consequence of
neutrino-Earth interactions is that the number of events detected
during the day (when there are no neutrino-Earth interactions) can be
statistically different from the number of events detected during the
night. The Super-Kamiokande experiment has already presented
experimental data which seem to slightly prefer a nonzero day-night
asymmetry, even though the result is not yet statistically significant
\cite{SuperK_daynight} (the most recent result is
$A_{DN}=0.065\pm0.031\pm0.013$).

In this section we review the electron neutrino regeneration effect in
the Earth and how it modifies the solar neutrino data. We also present
the expected day-night asymmetry for $^7$Be neutrinos at the KamLAND
and Borexino sites.

\subsection{The Day-Night Effect}

If neutrinos have mass, it is very likely that, similar to what
happens in the quark sector, neutrino mass eigenstates are different
from neutrino weak eigenstates. Assuming that only two neutrino states
mix, the relation between mass eigenstates and flavor eigenstates is
simply given by
\begin{eqnarray}
\label{masseigen}
|\nu_1\rangle=\cos\theta|\nu_e\rangle-\sin\theta|\nu_{\mu}\rangle,
\nonumber \\
|\nu_2\rangle=\sin\theta|\nu_e\rangle+\cos\theta|\nu_{\mu}\rangle, 
\end{eqnarray} 
where $\theta$ is the vacuum mixing angle, $|\nu_1\rangle$ and
$|\nu_2\rangle$ are the mass eigenstates with masses $m_1$ and $m_2$,
respectively, and $\nu_e\leftrightarrow \nu_{\mu}$ mixing is
considered. The mass-squared difference is defined as $\Delta
m^2\equiv m_2^2-m_1^2$.  We are interested in the range of parameters
that encompasses all physically different situations. First, observe
that Eq.~(\ref{masseigen}) is invariant under $\theta \rightarrow
\theta+\pi$, $\nu_e \rightarrow -\nu_e$, $\nu_\mu \rightarrow
-\nu_\mu$, {\it i.e.} $\theta\in [-\pi/2,\pi/2]$ and $\theta\in
[\pi/2,3\pi/2]$ are physically equivalent. Next, note that it is also
invariant under $\theta \rightarrow -\theta$, $\nu_\mu \rightarrow
-\nu_\mu$, $\nu_2 \rightarrow -\nu_2$, hence it is sufficient to only
consider $\theta\in [0,\pi/2]$.  Finally, it can also be made
invariant under $\theta \rightarrow \pi/2 - \theta$, $\nu_\mu
\rightarrow -\nu_\mu$ by relabeling the mass eigenstates $\nu_1
\leftrightarrow \nu_2$, {\it i.e.} $\Delta m^2 \rightarrow -\Delta
m^2$. Thus, all physically different situations are obtained if
$0\leq\sin^2\theta\leq1$ and $\Delta m^2$ is positive, or
$0\leq\sin^2\theta\leq 1/2$ and $\Delta m^2$ can have either sign. In
what follows, we will use the first parametrization ($\Delta m^2>0$),
unless otherwise noted.

$^7$Be neutrinos reach the Earth as an incoherent mixture of
$|\nu_1\rangle$ and $|\nu_2\rangle$ \cite{incoherent} (see also
\cite{incoherent2,Guth_Randall}), with probabilities $P_1$ and $P_2=1-P_1$ as
long as $\Delta m^2\gtrsim 10^{-8}$~eV$^2$.  $P_1$ is given in
Eq.~(\ref{P1}) in terms of the jumping probability $P_c$ 
and its value depends on the details of the
neutrino production and propagation inside the Sun, as presented in
Appendix~\ref{negativedeltam2}.  The probability $P_{ee}$ of detecting
a $\nu_e$ on the Earth is given by
\begin{equation}
\label{pee1}
P_{ee}= P_1 P_{1e} + (1-P_1) P_{2e}\ , 
\end{equation}           
where $P_{ie}$ is the probability that $\nu_1$ ($\nu_2$) is detected
as a $\nu_e$ for $i=1$~(2). Because $P_{1e}+P_{2e}=1$ (always,
independent of matter effects, because of the unitarity of the
Hamiltonian), one can rewrite
Eq.~(\ref{pee1})  
\begin{equation}
\label{pee2}
P_{ee}= P_1 + (1-2 P_1)P_{2e}\ .
\end{equation}
In the case of neutrinos detected during the day, $P_{2e}=\sin^2\theta$
(the vacuum result), while for neutrinos that traverse the Earth
$P_{2e}=P_{2e}^{E}$ must be calculated numerically, and depends on the
density profile of the Earth and the latitude of the location where
the neutrinos are to be detected. One should also remember that muon
or tau neutrinos still interact in the detector through neutral
currents, although the even rate is down by a factor of
$R\simeq 0.2$ compared to electron neutrinos. The day-night asymmetry
($A_{DN}\equiv$ (events detected during the night minus events detected during
the day)/(total)) is, therefore,
\begin{equation}
\label{dna}
A_{DN}=\frac{(1-2P_1)(P_{2e}^E-\sin^2\theta)(1-R)}
{(2P_1 + (1-2 P_1)(P_{2e}+\sin^2\theta))(1-R)+2R} \, .
\end{equation}
It is important to note that $A_{DN}$ 
does not have to vanish, as used to be the general lore in the past, when
$\sin^2\theta=1/2$ (maximum mixing), as was clearly shown in
\cite{Guth_Randall}. $A_{DN}$ does vanish, of course, when $P_1=1/2$ (a
fifty-fifty mixture of mass eigenstates reaches the Earth).

It is interesting to note that, in the past, $A_{DN}$ was always
computed assuming that $\sin^2\theta\leq 1/2$. However, it is
perfectly acceptable to have $\sin^2\theta > 1/2$, when the {\it
heavy} mass eigenstate ($\nu_2$) is predominantly $\nu_e$. While in 
the case of vacuum oscillations physical results depend only on $\sin^2
2\theta$, in the case of neutrino-matter interactions  
$\sin^2\theta > 1/2$ leads to physically different results. Using
$\sin^2 2\theta$ as a parameter in the latter case can be
misleading, as $0\leq\sin^2 2\theta\leq 1$ does not cover all
physically distinct possibilities. Similar
to what was pointed out in \cite{Guth_Randall} for the transition
between $\sin^2\theta < 1/2$ to $\sin^2\theta = 1/2$, we will show
that for the entire range of $0\leq\sin^2\theta\leq1$ the behavior of
$A_{DN}$ is smooth. In Appendix~\ref{negativedeltam2} we explain in
detail how to extend the expression for $P_1$ to the case
$\sin^2\theta>1/2$.

\subsection{The Day-Night Asymmetry at 36$^{\rm o}$ and 42$^{\rm o}$ North}
\label{sect:dn_asym}

We numerically compute the value of $P_{2e}^E$ and $A_{DN}$ for $^7$Be
neutrinos at KamLAND (latitude = 36.4$^{\rm o}$ north) and Borexino 
(latitude = 42.4$^{\rm o}$ north).  
We assume a radially symmetric exponential profile for
the electron number density inside the Sun, and use the analytic
expression for the survival probability of neutrinos produced in the
Sun's core derived in \cite{Petcov}, as presented in
Appendix~\ref{negativedeltam2}. We appropriately integrate over the
$^7$Be neutrino production region inside the Sun, using the results of
the SSM \cite{SSM}, conveniently tabulated in \cite{bahcall_www}.  

We use a radially symmetric profile for the Earth's electron number
density, given in \cite{earth_profile}, and the zenith angle exposure
function for the appropriate latitude, which was obtained from
\cite{bahcall_www}. For a plot of the electron number
density profile in the Earth see Fig. 2 in \cite{day_night} and for
the zenith angle exposure function see the upper left-hand corner of
Fig. 5 in \cite{day_night}. The model predicts that the electron
number density in the Earth's mantle varies in the range 2.1 to 2.7
moles/cm$^3$, while in the outer core the electron number density is
significantly greater (4.6 to 5.6 moles/cm$^3$). Because of the
latitude of Borexino and KamLAND, the solar neutrinos detected at
these experiments will not travel through the inner core. 

\begin{figure} [htbp]
\centerline{
  \psfig{file=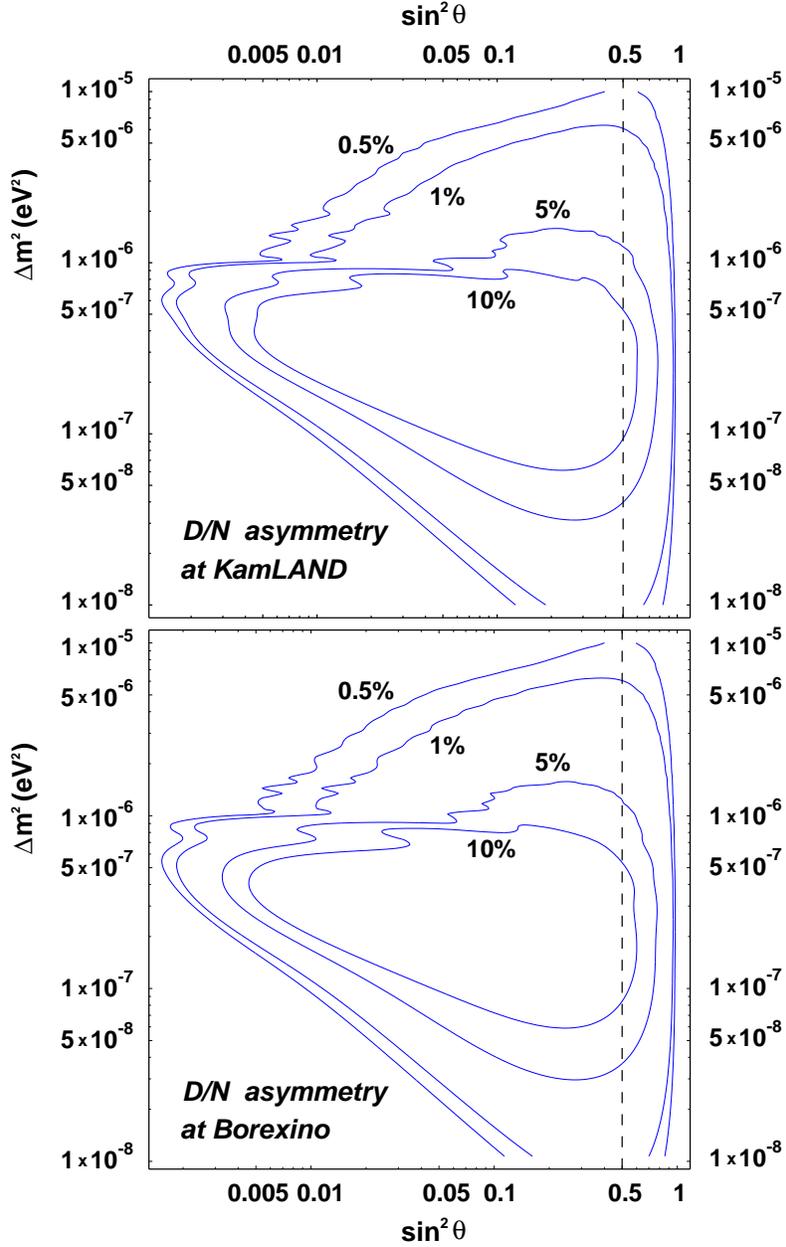,width=0.76\textwidth}
}

\caption{Constant day-night asymmetry contours ($10\%, 5\%,
  1\%, 0.5\%$) in the ($\sin^2\theta,\Delta m^2$)-plane for $^7$Be
  neutrinos at the KamLAND and Borexino sites.  The vertical dashed
  line indicates $\sin^2\theta=1/2$, where the neutrino vacuum mixing
  is maximal.}
\label{asym_plot}
\end{figure}
\begin{figure}
  \begin{center}
      \psfig{file=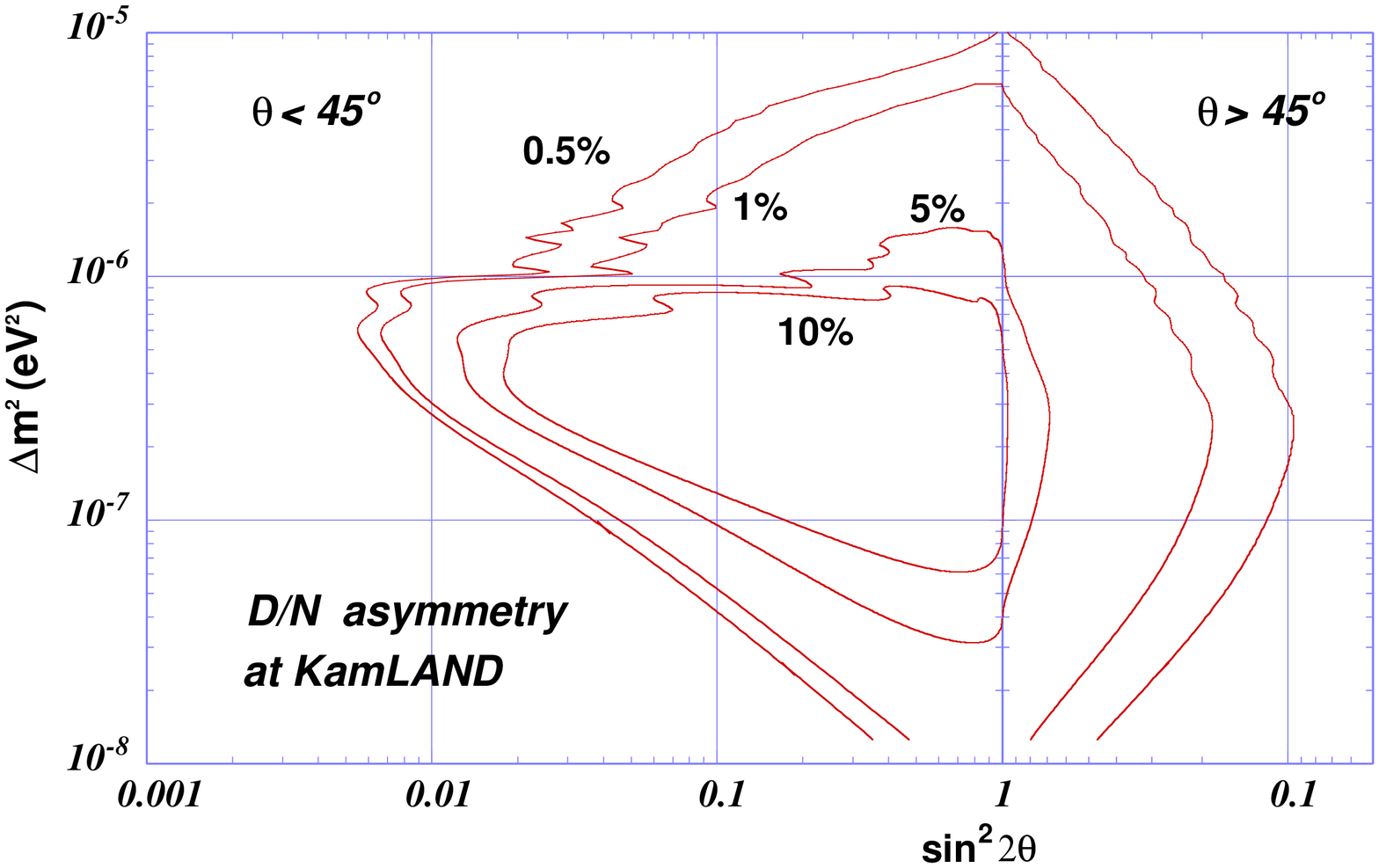,width=0.9\textwidth}
      \psfig{file=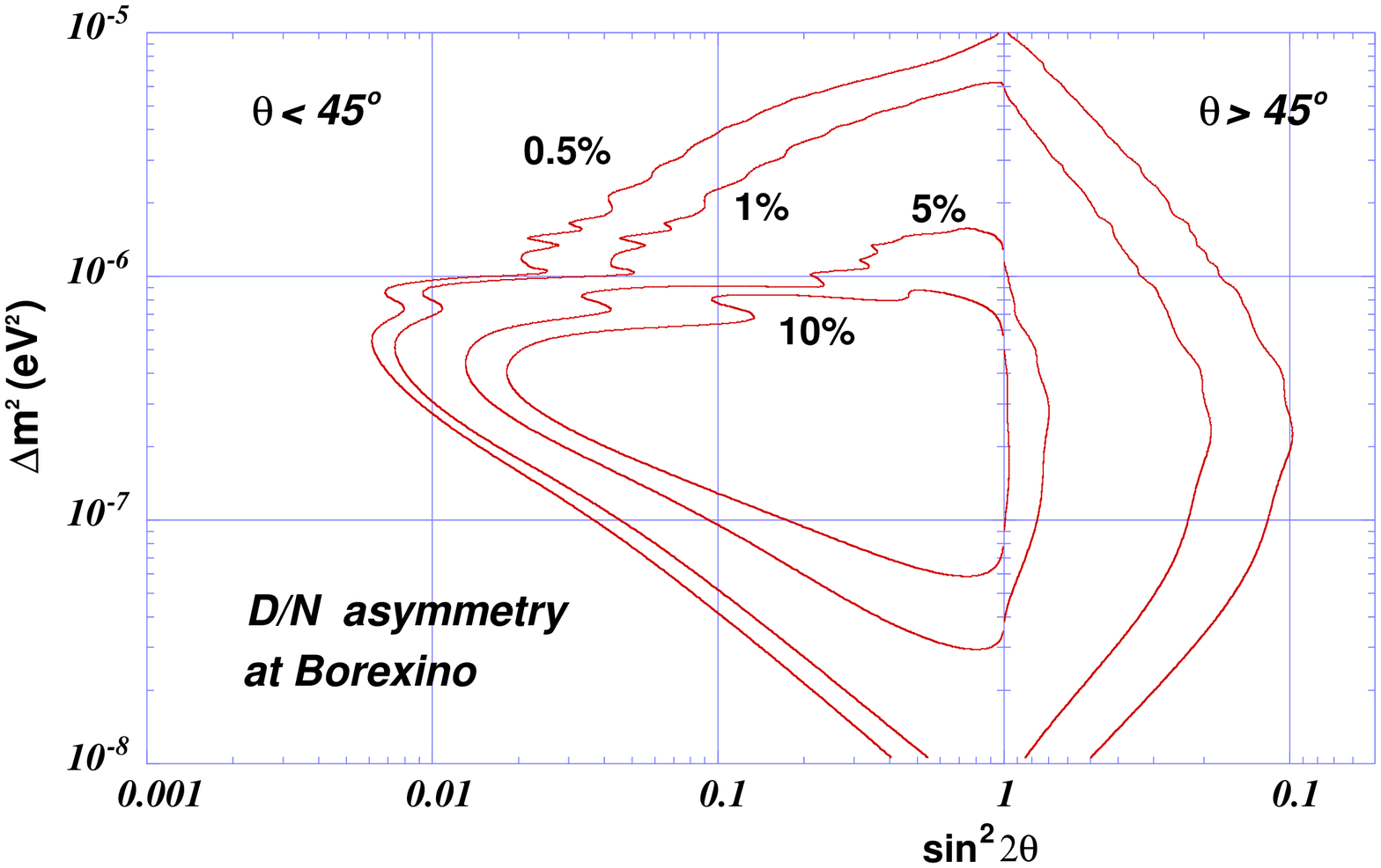,width=0.9\textwidth}
  \end{center}
\caption{Constant day-night asymmetry contours ($10\%, 5\%,
  1\%, 0.5\%$) in the
  ($\sin^22\theta,\Delta m^2$)-plane for $^7$Be neutrinos at the
  KamLAND and Borexino sites. The right side of the plot, with
  decreasing scale, can also be thought of as $\Delta m^2<0$,
  $\theta<45^{\rm 0}$.}
\label{asym_plot_usual}
\end{figure}
Fig.~\ref{asym_plot} depicts the constant day-night asymmetry contours
for $^7$Be neutrinos\footnote{We only assume $\nu_e$ oscillations into
other active neutrino species.} at KamLAND and Borexino.  It is
important to note that, unlike conventionally done in the literature,
the $x$-axis here is $\sin^2\theta$, not $\sin^22\theta$. To facilitate
comparison with earlier results, we also depict the same information in
the ($\Delta m^2,\sin^22\theta$) plane in Fig.~\ref{asym_plot_usual}, 
where once again we vary the mixing angle in its entire physical range
$0\leq\theta\leq\pi/2$.

As Fig.~\ref{asym_plot} demonstrates, the asymmetry contours smoothly
extend into the $\sin^2 \theta > 0.5$ half of the parameter space.
One can see that in that region the day-night asymmetry is non-zero
and may, in fact, be quite large. This kind of behavior had already been seen
in \cite{FLM}, for day-night asymmetry contours at SuperKamiokande (see Fig.~11
in \cite{FLM}).
This is to be contrasted with conventional analyses, which choose
axes as in Fig.~\ref{asym_plot_usual}, but only show the
$0\leq\theta\leq\pi/4$ half of the parameter space. As a result,
contours there seem to abruptly terminate at maximal mixing. 

It is also easy to see from our plots that, with the choice of
variables as in Fig.~\ref{asym_plot}, there is nothing special about
maximal mixing. This point is somewhat obscured in the ($\Delta
m^2,\sin^22\theta$) plane, where it seems that the slope of the
contours abruptly changes around $\sin^22\theta=1$. The reason for
this is that the Jacobian of the transformation from $\sin^2\theta$ to
$\sin^22\theta$,
\begin{equation}
  \label{eq:jacobian}
  \frac{d (\sin 2\theta)}{d (\sin\theta)}=2\frac{\cos 2\theta}{\cos\theta},
\end{equation}
vanishes at maximal mixing $\theta=\pi/4$.  It can be argued, therefore, that
$\sin^2\theta$ represents a more natural parametrization. From here on
we will always use $\sin^2\theta$ as a parameter.\footnote{If one
  wishes to keep the symmetry between $\theta<\pi/4$ and
  $\theta>\pi/4$ for vacuum oscillations while avoiding the singular
  Jacobian, the best choice for the horizontal axis would be
  $\tan\theta$ in log scale, as was done in \cite{FLM} in the context of 
  three-flavor oscillations.   }

The day-night asymmetry for $\theta=\pi/4$ is in general non-zero and,
indeed, can be larger than 10\%. Our analysis, thus, is in
complete agreement with the findings of \cite{Guth_Randall} and
extends them to the other half of the parameter space.
Note that constant
day-night asymmetry contours do close as $\sin^2\theta\rightarrow 1$. 
This is expected, because in that limit, just like for 
$\sin^2\theta\rightarrow 0$, there is no neutrino
mixing, and so $P_{ee}$ goes trivially to 1 and $A_{DN}$ vanishes.

Almost all other features of the contours in Figs.~\ref{asym_plot} and
\ref{asym_plot_usual} can also be understood analytically. Several
physical effects are involved in shaping up the contours. In the low
$\Delta m^2$ region the oscillation length in the Earth is comparable
to the size of the Earth, independent of the value of $\Delta
m^2$. This can be understood very easily in the approximation that the
Earth's electron density is uniform. In that case the neutrino
oscillation length is given by
\begin{eqnarray}
L_{osc}=\pi
\left[\left(\frac{\Delta
    m^2}{2E_{\nu}}\right)^2+(\sqrt{2}G_{F}N_{e})^2
  -2\left(\frac{\Delta m^2}{2E_{\nu}}\right)
    \sqrt{2}G_{F}N_{e}\cos2\theta\right]^{-1/2},
\end{eqnarray}
or numerically
\begin{eqnarray}
  \label{eq:osc_length}
L_{osc}&=&
10.7\times 10^4\: {\rm km}\left[\left(\frac{\Delta
    m^2}{10^{-7}\: {\rm eV}^2}\right)^2+
\left(1.3\frac{N_{e}}{1\:{\rm mole/cm^3}}\right)^2\right.\nonumber\\
 &-&\left.2\left(\frac{\Delta m^2}{10^{-7}\: {\rm eV}^2}\right)
    \left(1.3
\frac{N_{e}}{1\:{\rm mole/cm^3}}\right)\cos2\theta\right]^{-1/2},
\end{eqnarray}
and, for $\Delta m^2/(2E_{\nu})\ll \sqrt{2}G_{F}N_{e}$,
$L_{osc}\rightarrow 8.2\cdot10^3{\:\rm km}\times(1\:{\rm
mole/cm^3}/N_{e})$. 

For very small $\Delta m^2$ the asymmetry vanishes for two reasons.
First, the MSW transition inside
the Sun becomes non-adiabatic.  For $\Delta m^2 \ll 10^{-5}$~eV$^2$,
$\theta_M \simeq \pi/2$ (Eq.~(\ref{theta_matter})) in the Sun's core
and $P_1 \simeq P_c$ (Eq.~(\ref{P1})).
As the value of the jumping probability
$P_c$ changes from 0 to $\cos^2\theta$ it passes through 1/2 (for
$\theta<\pi/2$) where $A_{DN}$ vanishes, according to
Eq.~(\ref{dna}).  As can be
deduced from Eq.~(\ref{Pc}), the contours 
of constant jumping probability $P_c$ are approximately described by
$\Delta m^2 \sin^2\theta=$ constant, provided $\sin^2 \theta \ll 1$
and $\Delta m^2 \gg 10^{-9}$~eV$^2$. Second, the mixing angle in
the Earth becomes close to $\pi/2$ and no regeneration takes place in
that limit (see also Eq.~(\ref{eq:Pav_uniform}) below, where $\theta_M
\rightarrow \pi/2$ gives $P_{2e}^{av} \rightarrow \sin^2 \theta$). Below the 
line $P_1=1/2$ the asymmetry is negative and very small. 

In the region $\Delta m^2 \gtrsim 3\times10^{-6}$~eV$^2$ neutrinos
undergo many oscillations inside the Earth, as can be seen from
Eq.~(\ref{eq:osc_length}). The relevant quantity in this case is the
average survival probability, obtained after integrating over the
zenith angle.  One can understand the shape of the asymmetry contours
in this region by, once again, approximating the electron number density
in the Earth by a constant value. In this model, it is easy to show that,
if a state $|\nu_i\rangle$ enters from vacuum into the Earth, the
average survival probability inside the Earth is
\begin{eqnarray}
  \label{eq:Pav_uniform}
  \begin{array}{cc}
  P_{2e}^{av}=\sin^2 \theta_M + \sin^2(\theta-\theta_M) \cos2\theta_M, &
   \;\;|\nu_i\rangle=|\nu_2\rangle, \\
  P_{1e}^{av}=\sin^2 \theta_M + \cos^2(\theta-\theta_M) \cos2\theta_M, &
   \;\;|\nu_i\rangle=|\nu_1\rangle. \\
  \end{array}  
\end{eqnarray}
Here $\theta$ is the mixing angle in vacuum and $\theta_M$ is the mixing
angle inside the Earth (see Eq.~(\ref{theta_matter})). Obviously,
$P_{1e}^{av}+ P_{2e}^{av}=1$. Using these expressions, one can
compute the day-night asymmetry for this simplified model:
\begin{equation}
  \label{eq:asym_uniform}
  A = \frac{P_N+(1-P_N)R-P_D-(1-P_D)R}{P_N+(1-P_N)R+P_D+(1-P_D)R},
\end{equation}
where
\begin{eqnarray*}
  P_D &=& \sin^2\theta_\odot((1-P_c)\sin^2\theta+P_c\cos^2\theta)+ \\ 
  &&\cos^2\theta_\odot((1-P_c)\cos^2\theta+P_c\sin^2\theta),\\
  P_N &=& \sin^2\theta_\odot((1-P_c) P_{2e}^{av}+P_c P_{1e}^{av})+ \\
  &&\cos^2\theta_\odot((1-P_c)P_{1e}^{av}+P_c P_{2e}^{av}).
\end{eqnarray*}
$\theta_\odot$ denotes the mixing angle at the production region in
the core of the Sun, $P_c$ the jumping probability
(Eq.~(\ref{Pc})), and $R$ is a contribution of $\nu_{\mu,\tau}$
interacting through the neutral current interactions in the detector.
We found that for $N_e\sim3-4$~moles/cm$^3$ the contours of constant
$A$ are in good agreement with the day-night asymmetry contours in
Fig.~\ref{asym_plot} for $\Delta m^2\gtrsim 3\times10^{-6}$~eV$^2$.

Using this simple model we can explain the behavior of the asymmetry
contours in the large $\Delta m^2$ region. For
example, according to Fig.~\ref{asym_plot}, as $\sin^2 \theta$
decreases for fixed $\Delta m^2$, the value of the asymmetry goes
down. This happens because, while the difference in the numerator of
Eq.~(\ref{eq:asym_uniform}) goes to zero, the denominator approaches a
constant value due to the non-vanishing neutral current
contribution. Notice that in a real experiment, in addition to the
neutral current contribution, there will be a term proportional to the
rate of background events, further decreasing the sensitivity. Thus,
using asymmetry contours in this region to read off the sensitivity
can be misleading. This would be even more obvious in the case of
oscillations to a sterile neutrino. We will return to this issue in
the next section.

Even more subtle features can be understood within this model. For
instance, we found that the slight change of the slope seen for the
0.5\% contour around $\sin^2 \theta\sim 0.04$ is due to the significant 
deviation of the value of $\theta_\odot$ from $\pi/2$ in that
region. 

Finally, in the region $\Delta m^2\sim10^{-6}$~eV$^2$ the regeneration 
efficiency exhibits a very strong zenith angle dependence. Because the
magnitudes 
of $\Delta  m^2/(2E_{\nu})$ and $\sqrt{2}G_{F}N_{e}$  in the core are
almost equal, the mixing in the core is almost maximal
($\theta_M\sim\pi/4$, see Eq.~(\ref{theta_matter})), while in the mantle
it is small ($\theta_M\sim \pi/2$). As a result, for neutrinos traveling
through the outer core the conversion into $\nu_e$ is much more
efficient than for ones going only through the mantle. The oscillations do not
average out completely in this case, resulting in the presence of
several wiggles. We have explicitly checked that these
wiggles are not washed out by the effect of the finite width of the
$^7$Be line \cite{lineprfl}.

Our results for $\theta<\pi/4$ agree qualitatively with the results
presented in 
\cite{day_night} for the Borexino site. The agreement is not complete, 
however. For instance, the contours in \cite{day_night} do not exhibit 
any wiggles in the range $\Delta m^2\sim10^{-6}$~eV$^2$.

\setcounter{equation}{0}
\setcounter{footnote}{0} 
\section{The Neutrino Regeneration Effect at KamLAND and Borexino}
\label{KamLAND}

In this section we study the sensitivity of the KamLAND and Borexino  
experiments to the day-night effect.

Borexino \cite{Borexino} is a dedicated $^7$Be solar neutrino
experiment. It is a large sphere containing ultrapure organic liquid
scintillator (300~t) and can detect the light emitted by recoil
electrons produced by elastic $\nu$-$e$ scattering. By looking in the 
appropriate recoil electron kinetic
energy window, it is possible to extract a very clean sample of events
induced by $^7$Be neutrinos, if the number of background events is
sufficiently small. Borexino  expects, in the absence of neutrino
oscillations, 53 neutrino induced events/day according to the
SSM, and 19 events/day induced by background 
(mainly radioactive impurities in the detector, see
\cite{seasonal,Borexino} for details).    

The KamLAND experiment, located in the site of the original
Kamiokande experiment, was initially designed as a reactor
neutrino experiment. Recently, however, the fact that 
KamLAND might be used as a solar neutrino experiment has become a plausible and
exciting possibility \cite{KamLAND}.

KamLAND is also a  very large sphere containing ultrapure liquid
scintillator (1~kt), and functions exactly like Borexino. The
outstanding issue to determine if KamLAND will study solar neutrinos
is if the background rates can be appropriately reduced. 
KamLAND expects, in the absence of neutrino
oscillations, 466 neutrino induced events/kt/day according to the
SSM, and 217 events/kt/day induced by background 
(mainly radioactive impurities in the detector, see
\cite{seasonal,KamLAND} for details). We will
consider a fiducial volume of 600~t, so that 280 (unoscillated)
signal events/day and 130 background events/day are expected. We
assume that the number of background events is constant in time.

We generate a histogram of the number of events expected in each of
the $N$ day and $N$ night bins for different values of ($\Delta
m^2,\sin^2\theta$). The number of events per year in the $i$-th bin is
\begin{equation}
\label{ni}
n_i\left(\frac{\rm events}{\rm year}\right)=365\left(\frac{\rm
  days}{\rm year}\right)(b_{\rm rate}+s_{\rm
  rate}(P_{ee}^i+(1-P_{ee}^i)R))
\left(\frac{\rm events}{\rm day}\right)f_i,
\end{equation}
where $s_{\rm rate}=280$ (53) events/day and $b_{\rm rate}=130$ (19)
events/day for KamLAND (Borexino), $P_{ee}^i$ is the electron neutrino 
survival probability in the $i$-th
bin, $R$ is the ratio of the $\nu_e$-$e$ to
$\nu_{\mu,\tau}$-$e$ elastic cross sections\footnote{In the case of 
  $\nu_e\leftrightarrow \nu_{\rm sterile}$ oscillations, $R=0$.} 
(see \cite{seasonal}, at KamLAND (Borexino) $R=0.214$ (0.213))
and $f_i$=(size $i$-th bin divided by the sum of the sizes of all
the bins), such that $\sum_i^{2N} f_i=1$. As an example, if there are 24 (12
day, 12 night) hour-bins,
$f_i=1/24$ for all $i$. In reality, we are interested in zenith angle
bins, and in order to the determine $f_i$, the exposure function
presented in \cite{day_night} is used. Note that we assume only
statistical uncertainties.  

$\chi^2$ is defined as
\begin{equation}
\label{chi2}
\chi^2=\sum_{i=1}^{N}\frac{\left(n_i^{\rm night}-n_i^{\rm day}\right)^2} 
{\left(\sqrt{n_i^{\rm night}}\right)^2+\left(\sqrt{n_i^{\rm day}}\right)^2}+N.
\end{equation}
The factor $N$ is included in the definition of $\chi^2$ in
order to take statistical fluctuations of the data into account. A
detailed explanation of the philosophy behind this procedure can be
found in \cite{seasonal}.

It is important to comment at this point that, in light of the
definition of $\chi^2$ (Eq.~(\ref{chi2})), the sensitivity of the
experiments to the Earth matter effect does not require any input from
the SSM, including the 
$^7$Be solar neutrino flux, or from a direct measurement of the
background rate. This is because we are comparing the night data with
the day data, and no other inputs are required. Our quantitative
results, however, depend on the expected number of signal and
background induced events, since these quantities are used as input
for the ``data'' sample. 

We will define the sensitivity of a given experiment to the Earth
matter effect by the value of $\chi^2$, computed according to
Eq.~(\ref{chi2}). The sensitivity defined in this way depends clearly
on $N$, the number of day and night bins, and on $f_i$ (see
Eq.~(\ref{ni})), or on the ``size of the bin''. With the real
experimental data, one will certainly consider many different types of
analyses in order to maximize the sensitivity of the data to the
neutrino regeneration in the Earth (options include computing moments
of the zenith angle distribution, Fourier decomposing the data,
maximum likelihood analysis, and others), but, since we analyze
thousands of ``data samples'' (one for each value of ($\Delta m^2,
\sin^2\theta$)), this simple $\chi^2$ approach will suffice.

We consider two options for the size of zenith angle bins. In one of
them, each bin has the same size, that is, the bins are equally spaced
({\it e.g.}\/ $0^{\rm o}-30^{\rm o}, 30^{\rm o}-60^{\rm o}, 60^{\rm
o}-90^{\rm o}$, etc). The other option is to choose the bin size such
that the distribution of the day data is uniform. 
It is worthwhile to comment that the latter scheme may be considered the
most natural one for KamLAND and Borexino, which are real time
experiments with no directional capability. In these experiments, it is
straightforward to organize the data into time bins, which then have
to be translated into zenith angle bins by associating the time of the
event with the position of the Sun in the sky.
 
\begin{figure} [htbp]
\centerline{
  \psfig{file=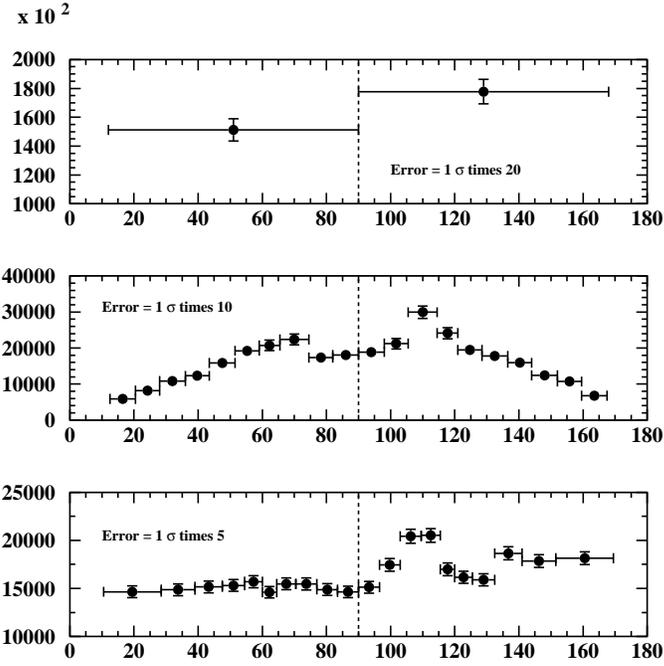,width=0.7\textwidth}
}
\caption{Different binning schemes, for $\Delta m^2=1.12\times10^{-7}$~eV$^2$,
  $\sin^2\theta=0.398$: (a) $N=1$~bin (the day-night asymmetry), (b)
  $N=10$~equally spaced zenith angle bins, and (c) $N=10$~``uniform'' 
  bins, where the day-time data is (roughly) uniformly distributed. The error
  bars contain statistical uncertainties only. We assume three years
  of KamLAND running.}
\label{sample_bins}
\end{figure}

Another issue to consider is the value of $N$ which optimizes the
sensitivity. It is clear that for $N=1$ (the day-night asymmetry case)
the statistical significance is enhanced for overall changes in the
number of events, but for larger $N$, one should be more sensitive to
distortions in the zenith angle distribution. Different binning
schemes of the ``data'' 
for $\Delta m^2=1.12\times10^{-7}$~eV$^2$, 
$\sin^2\theta=0.398$ and three years of KamLAND running
are depicted in Fig.~\ref{sample_bins}, for $N=1$, $N=10$ equally
spaced bins, and $N=10$ ``uniform'' bins.\footnote{The residual
non-uniformity seen in the figure is due to the fact that we used a
discrete table of values for the exposure function.}

\begin{figure} [htbp]
\centerline{
  \psfig{file=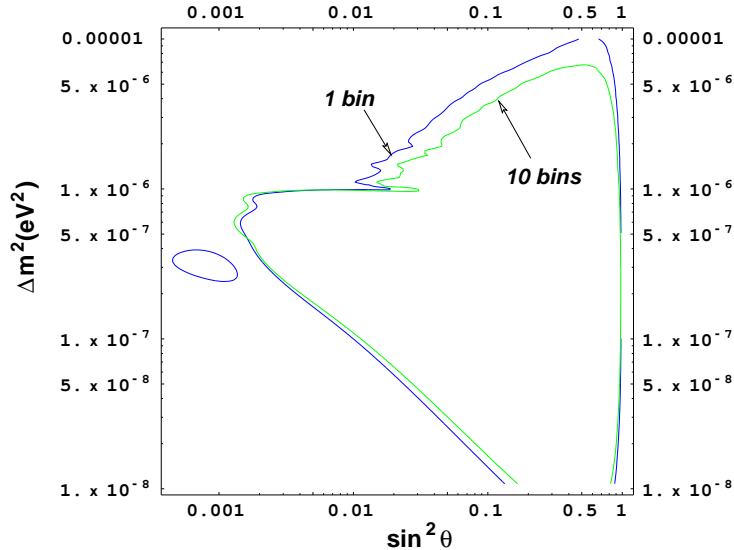,width=0.7\textwidth}
}
\caption{Comparison of the sensitivity reach of three years of KamLAND
  running with 1 bin and 10 uniform bins.}
\label{comparison}
\end{figure}

Fig.~\ref{comparison} shows a comparison of the sensitivity reach of
KamLAND after three years of running for two different
binning schemes, $N=1$ vs. $N=10$ ``uniform'' bins. The contours are
drawn at 95\% C.L.  One can easily see that for most of the parameter
space, the best sensitivity is reached with the $N=1$
case, while for a small region in the parameter space, when
$\sin^2\theta\lesssim 0.1$ and $\Delta m^2\sim 10^{-6}$~eV$^2$, the
$N=10$ scheme is more successful.
This result is consistent with the analysis of Section
\ref{sect:dn_asym}. As explained there, for $\Delta m^2\sim
10^{-6}$~eV$^2$ the data shows a large enhancement in the low zenith
angle bin, while little effect in other bins. At Borexino this effect
will be somewhat less pronounced because it is farther from the Equator.

One can see that the contours in Fig.~\ref{comparison} are similar
in shape to the day-night 
asymmetry contours of Section~\ref{sect:dn_asym}, but quantitatively
different. One important difference is that for 
$\Delta m^2\gtrsim10^{-6}$~eV$^2$ the $\chi^2$ contours do not extend as 
far in the low $\sin^2 \theta$ 
region as the asymmetry contours. While for low $\Delta m^2$ the 95\%
C.L. contour corresponds to the day-night asymmetry of roughly 0.5\%,
for $\Delta m^2\gtrsim10^{-6}$~eV$^2$ the corresponding value of the
day-night asymmetry is at least two times greater. This phenomenon was 
already mentioned in Section~\ref{sect:dn_asym}. The difference occurs
because the $\chi^2$ analysis includes, in addition to the neutral
current interactions, the constant background rate, thus eliminating the 
major shortcoming of the day-night asymmetry analysis.

\begin{figure} [htbp]
\centerline{
  \psfig{file=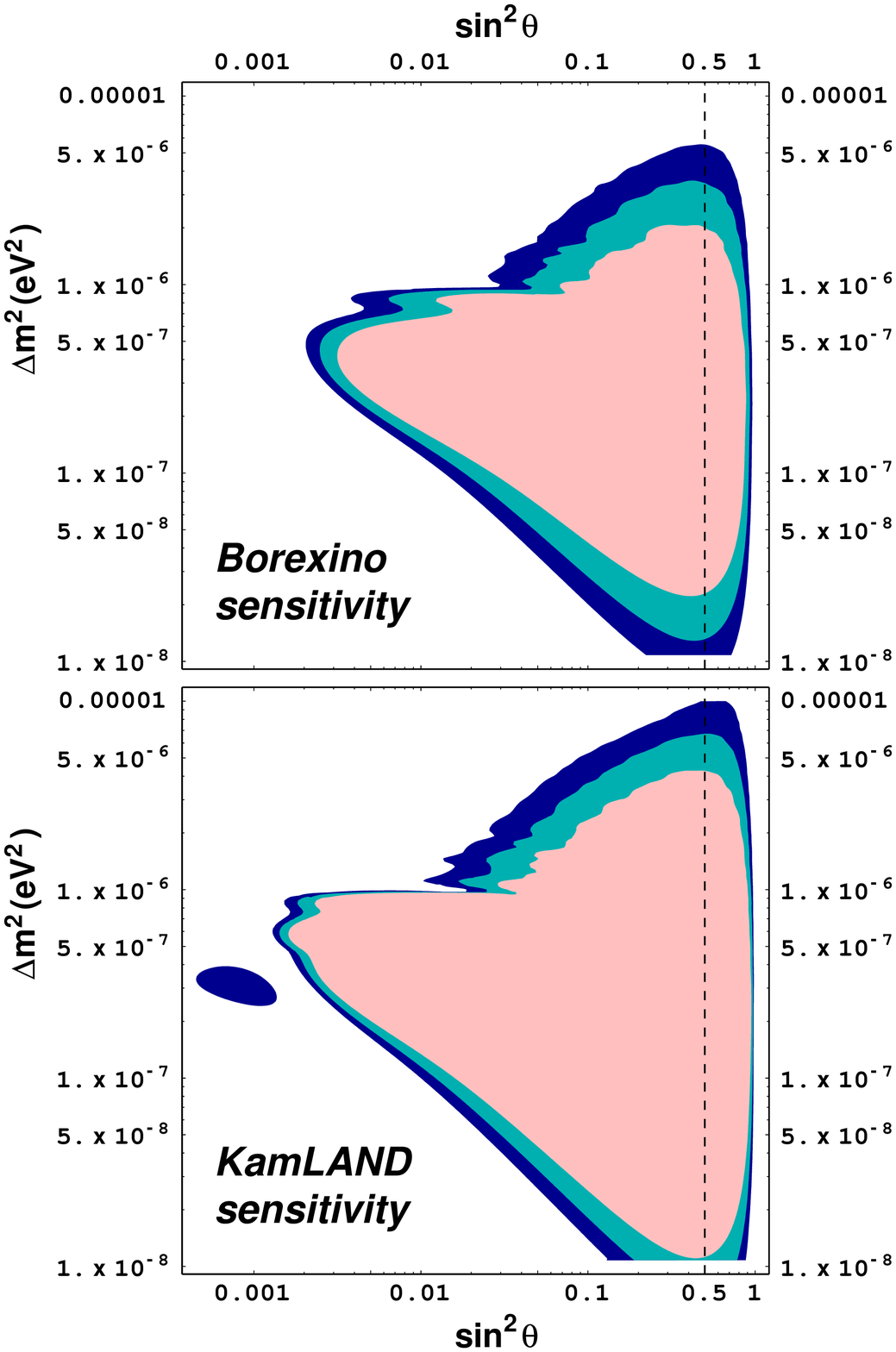,width=0.81\textwidth}
}

\caption{95\% (darkest), 3$\sigma$ (dark), and
5$\sigma$ (light) sensitivity confidence level (C.L.) contours for
three years of KamLAND running. The LOW solution, which extends from
$\Delta m^2\sim 3\times10^{-8}$~eV$^2$ to $\Delta m^2\sim
3\times10^{-7}$~eV$^2$ and has $\sin^2\theta\sim0.3-0.5$
\cite{rate_analysis}, is completely covered at more than $5\sigma$
C.L.}
\label{optimal}
\end{figure}

In order to present the final sensitivity reach of KamLAND and
Borexino, we combine
the confidence level contour obtained in the different types of
analyses, with different number of bins. Fig.~\ref{optimal} depicts
the ``optimal'' 95\%, 3$\sigma$, and
5$\sigma$ confidence level (C.L.) contours for the sensitivity of
three years of KamLAND and Borexino data to the day-night effect. The
confidence levels are optimized
by considering the union of same C.L. contours for all
values of $N$ and both binning schemes. The day-night asymmetry
provides the best sensitivity reach for most of the parameter space,
while the $N\approx 10$ uniform bins scheme at KamLAND increases the 
sensitivity for particular regions of the parameter space, as was
discussed earlier. 

Fig.~\ref{optimal} clearly demonstrates that in the case of the LOW
MSW solution to the solar neutrino puzzle, both KamLAND and Borexino
should be able to see a larger than 5$\sigma$ effect, while in the
case of the SMA no significant effect should be
detected.\footnote{KamLAND may also be sensitive to a very small
portion of the LMA solution.} Both experiments are sensitive to a
large portion of the parameter space which extends into $\theta>\pi/4$
region, where the heavy neutrino eigenstate is predominantly
$\nu_e$.

On the other hand, should no regeneration effect be observed, a large
portion of the parameter space, including the entire LOW region might
be excluded. The exclusion will require knowledge of the $^7$Be
neutrino flux, which can be measured, for example, by studying the
seasonal variation of the observed event rate \cite{seasonal}. If the
flux measured in this way turns out large and no day-night asymmetry
is observed, one will be able to exclude the LOW solution without
relying on the solar model. If, however, the measured flux is very small,
the exclusion will be solar model dependent.

Since the sensitivity of Borexino (KamLAND) to the day-night asymmetry
goes down to the 1.5\% (0.5\%) level, it is important to consider
systematic effects in this measurement.  It is, however, difficult to
anticipate systematic uncertainties in the absence of data.  We
instead looked at the measurement of day-night asymmetry at
SuperKamiokande \cite{SuperK_daynight}. The dominant systematic
uncertainty there is the possible asymmetry in the detector, giving
$\pm 0.6\%$.\footnote{Note that the talk in \cite{SuperK_daynight}
lists the systematic uncertainties in $D/N$ ratio, which are twice as
large as uncertainties in the asymmetry $(D-N)/(D+N) \approx
((D/N)-1)/2$.}  Because the recoil electrons from $^8$B neutrinos are
forward peaked, the day (night) time data are detected primarily by
the lower (upper) half the detector.  A small possible gain asymmetry
(\cite{SuperK_daynight} quotes 0.5\%) for different zenith angle bins
can result is a somewhat amplified difference in rates because the
energy spectrum is rather steep close to the threshold energy
(6.5~MeV).  The energy calibration was done using electron LINAC,
which at that time could shoot electrons only downwards and hence
could not study the asymmetry well enough.  The gain asymmetry is
known to exist from the study of decay electrons in the cosmic ray muon
data \cite{kasuga_thesis} as well as in spallation events
\cite{suzuki_private}.\footnote{The gain asymmetry is now accurately
measured using the $^{16}$N source calibration and will be reduced
dramatically \cite{svoboda_private}.}  We assume that this will not be
an important systematic effect for Borexino or KamLAND because the energy
deposit is basically isotropic (no directional capability) and hence
the asymmetry in the detector should not result in a systematic
effect in the day-night asymmetry.

The next largest systematic effect is the subtraction of background,
$\pm 0.2\%$.  If the background events are not completely isotropic,
the subtraction depends on the direction and results in a systematic
effect.  Again at Borexino or KamLAND, the lack of directional
correlation eliminates this systematic effect.  

If we naively drop these two dominant systematic effects, the size of
the total systematic uncertainty would be less than 0.1\%.  Of course,
the sources of background are very different at Borexino or KamLAND.
Possible differences in the temperature or Rn level between the day
and night times could introduce new systematic effects, while our
analysis assumed the same background level for day and night.  This
difference, however, can in principle be measured using the Bi-Po
coincidence.  Spallation background (such as $^{11}$C) should not
change between day and night.  

Additionally, the experiments will need to consider other effects,
such as the contribution of other neutrino sources or the uncertainty
in the electron number density profile of the Earth. (More on the
latter in the next section.) We also did not include in our analysis
the contribution of neutrinos produced in the CNO cycle, which is
about 10\% of that from the $^7$Be neutrinos.
Although we cannot accurately predict
the total systematic uncertainty at Borexino or KamLAND, we
nonetheless find it encouraging that the dominant uncertainties at
SuperKamiokande are unlikely to affect these experiments.

\setcounter{equation}{0}
\section{Measuring the Oscillation Parameters}
\label{measurement}

In this section, we discuss the possibility of measuring the value
of $\Delta m^2,\sin^2\theta$ in the advent of a large day-night
effect. In order to do this, data was simulated for $\Delta m^2=1.12\times
10^{-7}$~eV$^2$, $\sin^2\theta=0.398$, which is close to the LOW MSW 
solution to the solar neutrino puzzle \cite{rate_analysis}. For a plot of the
``data'' with different binning options, see Fig.~\ref{sample_bins}.   

In order to deal with the SSM solar neutrino flux and
the background event rate, we will conservatively ``measure''
both the background rate and the incoming solar neutrino flux by analyzing
the seasonal variation \cite{seasonal} of the {\it day-time data only}. This 
measurement procedure will be incorporated in a four parameter $\chi^2$
analysis (the parameters are $\Delta m^2$, $\sin^2\theta$, the solar
neutrino flux $s$, and the background rate $b$) of the data. Explicitly,
\begin{equation}
\chi^2(\Delta m^2,
\sin^2\theta, s, b)=\sum^{N}_{i=1}\frac{\left({\rm data}_i^{\rm
      night} - {\rm theo}_i^{\rm dn}\right)^2}{\left( \sqrt{{\rm
    data}_i^{\rm night}}\right)^2} + \sum^{M}_{j=1}\frac{\left({\rm
data}_j^{\rm day}-{\rm theo}_j^{\rm sea}\right)^2} {\left( \sqrt{{\rm
    data}_j^{\rm day}}\right)^2},
\end{equation}
where data$_i^{\rm night}$ is the night-time ``data'' binned into $N$
night bins (as described in Sec.~\ref{KamLAND}), data$_j^{\rm day}$ is
the day-time ``data'' binned into $M$ ``seasonal bins'' ({\it e.g.}\/
$j=1,2, \dots 12$~months) as described in
\cite{seasonal}. theo$_i^{\rm dn}$ is the prediction for the number of
evens in the $i$-th night bin, 
\begin{equation}
{\rm theo}_i^{\rm dn}= 365\left[ b+s(P_{ee, i}^{\rm night}+(1-P_{ee,
    i}^{\rm night})R)\right]f_i,
\end{equation}
similar to Eq.~(\ref{ni}). $b$ is the background rate in events per
day and $s$ is the number of events per day induced solar neutrinos
according to the SSM prediction for the solar neutrino flux.
Similarly, theo$_j^{\rm sea}$ is the prediction for the day-time flux in
the $j$-th seasonal bin (see \cite{seasonal}),  
\begin{equation}
{\rm theo}_j^{\rm
  sea}=\left[\int_{i-1}^{i}{\rm d}t 
    \left(b+s\frac{P_{ee}+(1-P_{ee})R}{(1-\epsilon\cos(2\pi t/{\rm
    year}))^2}\right) \right]g_j,
\end{equation}
where $g_j$ is the number of days in the $j$-th bin and $\epsilon=0.017$ is
the eccentricity of the Earth's orbit. 

\begin{figure} [ht]
\centerline{
 \psfig{file=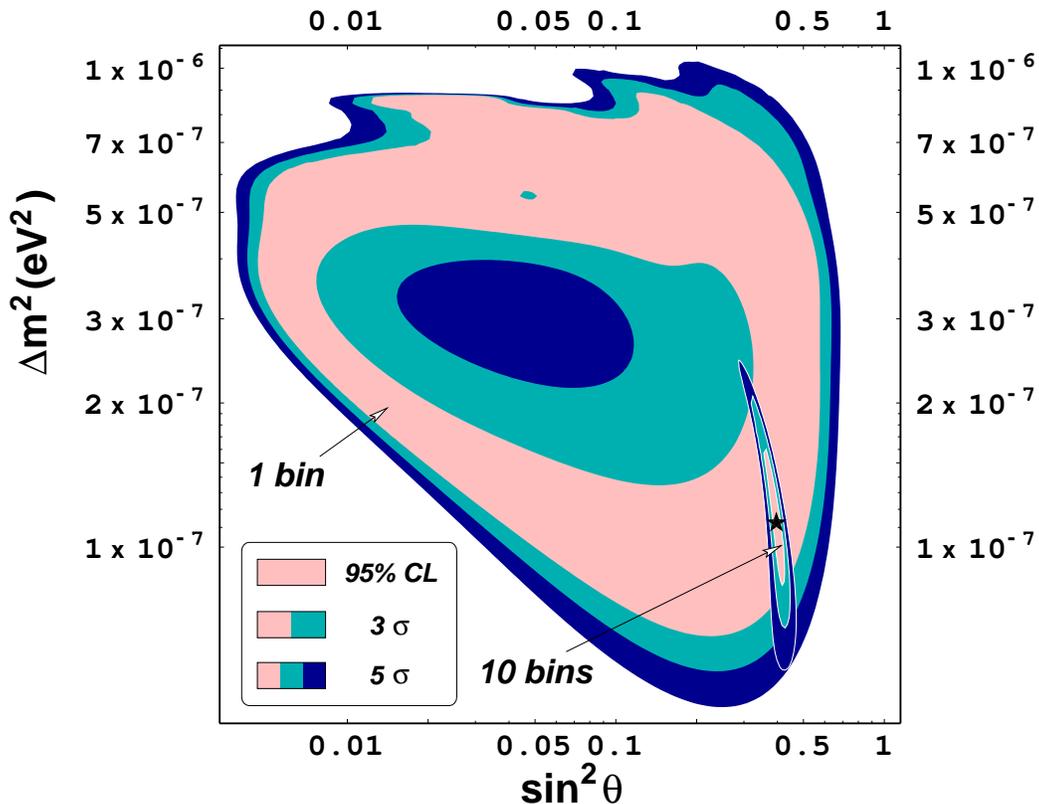,width=1.\textwidth}
}
\caption{Measured values of ($\Delta m^2,\sin^2\theta$) at KamLAND
  after three years of running. The data was generated for $\Delta
  m^2=1.12\times 10^{-7}$~eV$^2$, $\sin^2\theta=0.398$ (marked with
  the ``star''). The regions obtained by using one night bin and ten
  uniform night bins are shown.}
\label{measureplot}
\end{figure}

It is simple to minimize $\chi^2$ with respect to $s$ and $b$, given
that $\chi^2(s,b)$ is a quadratic function. The
minimization with respect to $\Delta m^2$ and $\sin^2\theta$ is done
numerically. Fig.~\ref{measureplot} 
depicts the extracted contours in the ($\Delta
m^2,\sin^2\theta$)-plane, in the case of 1 night bin
and 10 ``uniform'' night bins, respectively.

As Fig.~\ref{measureplot} demonstrates, in the case of 1 night bin,
one extracts values of $\Delta m^2$ and $\sin^2\theta$ which fall into
``rings'' which correspond roughly to $A_{DN}=A_{DN}^{\rm
real}\pm\Delta A_{DN}^{\rm real}$, where $A_{DN}^{\rm real}$ is the
value of the day-night asymmetry for the input value of $\Delta
m^2,\sin^2\theta$.  In the case of more than one uniform bin, the ring
degeneracy is broken, and a much more precise determination of the
oscillation parameters is possible.  This is expected, since for
$\Delta m^2$ in this range the regeneration effect in the Earth
exhibits a strong zenith angle dependence, as one can easily verified
by looking at Fig.~\ref{sample_bins}.

It is important to note that in the above analysis only statistical
uncertainties were included, while in a real experiment one definitely
will have to account for systematic effects as well. In particular,
one will need to address the uncertainty in the Earth model used in
the fit. In producing Fig.~\ref{measureplot} the same Earth model
\cite{earth_profile} was used in generating the ``data'' and in the
fit procedure. To understand the effect of using a ``wrong'' Earth
model, we have repeated the above analysis using different Earth models
in the fit. We found the results very encouraging.  Even in
the case when we used for the Earth profile a crude two-step model (a
uniform density in the mantle and a uniform density in the core), the
minimum of $\chi^2$ occurred at $\Delta m^2=2.5\times 10^{-7}$~eV$^2$, $\sin^2
\theta=0.24$, not far away from the true (input) value.  Moreover, the
$\chi^2$ value at the minimum was much larger than the case with the
``true'' model  ($\Delta \chi^2 = 183$ for 18
d.o.f.).  This means that in a real experiment one will be able to adjust
the Earth's model to achieve a better fit to the data.  Because of the
steep rise in $\chi^2$ value as the Earth model is varied, the
resulting $\chi^2$ contours in the $(\Delta m^2, \sin^2 \theta)$
parameter space should not be significantly larger than the ones
presented here, where the Earth model is not varied.  As a byproduct of the
measurement of the neutrino oscillation parameters, it might be
possible to use the regeneration data to study the interior of the
Earth!

\setcounter{equation}{0}
\section{Conclusions}
\label{conclusion}

We have studied the effect of the Earth matter on $^7$Be solar neutrinos.
We made use of an enlarged parameter space $0\leq\theta\leq\pi/2$
and presented the sensitivity reach of the KamLAND and Borexino
experiments in this space. Our results show that both experiments will
be sensitive to the Earth regeneration effect in a large region which
extends into the traditionally neglected $\theta>\pi/4$ part of the
parameter space. In particular, for the LOW solution one expects to
see a greater than $5\sigma$ effect. On the other hand, both
experiments will see no day-night effect for the SMA solution and
virtually no effect for the LMA solution.

If the experiments see a large Earth regeneration effect, it will be a
powerful ``smoking gun'' signature of neutrino
oscillations. Furthermore, as we have demonstrated, the results of the
experiments can be used to measure the oscillation parameters. By
studying the full zenith angle distribution, rather than the usual
day-night asymmetry information, one might be able, in the case of the LOW
solution, to perform a spectacular measurement of the parameters. In
addition, it might be possible to use the zenith angle information to
learn about the Earth electron density profile.

If, on the other hand, no Earth regeneration effect is detected, by
combining this information with the flux measurement from seasonal
variation of the event rate \cite{seasonal}, a large portion of the
parameter space can be excluded. If the measured value of the $^7$Be
neutrino flux is large, the exclusion will be independent of a specific
solar model.

Both the measurement of the oscillation parameters and the exclusion
will require a thorough understanding of the systematic
uncertainties. We have commented on some possible sources of such
uncertainties in this paper.

Overall, Borexino and KamLAND will provide crucial information about
the solar neutrinos. Not only will the experiments measure the flux of
the $^7$Be solar neutrinos, but they will also be able to establish or
exclude, without relying on solar models, the LOW solution based on
the Earth regeneration effect and the vacuum oscillation solution based on
the observed seasonal variation of the event rate. Together with 
results from SuperKamiokande, SNO, and the KamLAND reactor neutrino
experiment, this information can be used to finally unravel the
30-year-old solar neutrino puzzle.

\section*{Acknowledgements}
We thank Eligio Lisi for pointing out to us Ref.~\cite{FLM}, 
which considers, in the context of three-flavor oscillations, the 
$\theta>\pi/4$ case. 
This work was supported in part by the Director, Office of Science,
Office of High Energy and Nuclear Physics, Division of High Energy
Physics of the U.S. Department of Energy under Contract
DE-AC03-76SF00098 and in part by the National Science Foundation under
grant PHY-95-14797. HM thanks the Institute for Nuclear Theory at the
University of Washington for its hospitality and the Department of
Energy for partial support during the completion of this work. HM also
was supported by the Alfred P. Sloan Foundation and AdG by CNPq
(Brazil).

\appendix

\setcounter{equation}{0}
\section{Matter Oscillations and No Level Crossing}
\label{negativedeltam2}

In this appendix, we discuss the  survival probability
of solar electron neutrinos outside the Sun, in particular the case
of no level
crossing, {\it i.e.,}\/ when $\Delta m^2 \cos2\theta<0$ in the language
of the two neutrino mixing scenario.

In the literature, matter effects in the Sun are always considered
when there is ``level crossing'' inside of the Sun, {\it i.e.,}\/ when
the light neutrino is predominantly of the electron type, and, due
to neutrino-electron interactions, when the instantaneous Hamiltonian
eigenstate with the largest eigenvalue is predominantly of the
electron type in the Sun's core. The other case, when the {\it heavy}
neutrino is predominantly of the electron type, has not been studied
in the literature in the context of two neutrino oscillations. The authors of 
\cite{FLM}, however, have considered this possibility in the context of 
three-flavor oscillations. 

The reason for this apparent neglect is simple, and will become clear
as our results are presented. What happens is that, in the case of no
level crossing, the electron neutrino survival probability is always
bigger than 1/2, and therefore it seems that this scenario is not
relevant to the solar neutrino puzzle. This, however, may
not be the case \cite{new}. 

Before presenting the expressions for the electron neutrino survival
probability outside of the Sun, it is necessary to clearly define the
neutrino eigenstates and mixing angles. The neutrino mass eigenstates
are defined in Eq.~(\ref{masseigen}), and the notation introduced
in Sec.~\ref{dn_effect} will be used. In what follows our convention
is $\Delta m^2>0$ and $0\leq\sin^2\theta\leq1$.

Inside the Sun the Hamiltonian has the form
\begin{eqnarray}
  \label{eq:massmatrix}
\lefteqn{H=p_{\nu}+\frac{m^2_{\rm sum}}{4 p_{\nu}}
-\sqrt{2}G_F N_n(r)/2}\nonumber\\
 && +\left(
    \begin{array}{cc}
      -\frac{\Delta m^2}{4 p_{\nu}}\cos 2\theta + \sqrt{2}G_F N_e(r)&
      \frac{\Delta m^2}{4 p_{\nu}}\sin 2\theta \\
      \frac{\Delta m^2}{4 p_{\nu}}\sin 2\theta &
      \frac{\Delta m^2}{4 p_{\nu}}\cos 2\theta  \\
    \end{array}
  \right),
\end{eqnarray}
were $p_{\nu}$ is the solar neutrino momentum, $N_{e}(r)$ ($N_{n}(r)$)
is the electron (neutron) number density at a distance $r$ from the
Sun's core, $m^2_{\rm sum}\equiv m_2^2+m_1^2$ and $\Delta
m^2\equiv m_2^2-m_1^2$, $m_2^2$ and $m_1^2$ being the mass eigenvalues
in vacuum.  The eigenvalues of the instantaneous Hamiltonian are
\begin{eqnarray}
\lambda_{\pm}(r)&=&p_{\nu}+
\frac{m_{\rm sum}^2}{4p_{\nu}}+\frac{\sqrt{2}G_F (N_e(r)-N_n(r))}{2}
\pm\frac{1}{2}\left[\left(\frac{\Delta m^2}{2p_{\nu}}\right)^2\right.
\nonumber \\ 
&+&\left.\left(\sqrt{2}G_{F}N_{e}(r)\right)^2-\frac{\Delta
    m^2}{p_{\nu}}\sqrt{2}G_{F}N_{e}(r)\cos2\theta\right]^{1/2}.
\end{eqnarray}
For the study of neutrino oscillations, terms common to both states are 
irrelevant, and the first three terms can be dropped.
One can also safely replace $p_{\nu}$ by $E_{\nu}$ in
 the remainder. The instantaneous Hamiltonian eigenstates in terms of
 flavor eigenstates are
\begin{eqnarray}
|\nu_-(r)\rangle=\cos\theta_M(r)|\nu_e\rangle-\sin\theta_M(r)|\nu_{\mu}\rangle, \\
|\nu_+(r)\rangle=\sin\theta_M(r)|\nu_e\rangle+\cos\theta_M(r)|\nu_{\mu}\rangle.
\end{eqnarray}
Here 
$\theta_M(r)$ is the matter mixing angle, given by
\begin{eqnarray}
\label{theta_matter}
\lefteqn{\cos2\theta_M(r)=}\nonumber\\
&&\frac{\Delta
    m^2\cos2\theta-2E_{\nu}\sqrt{2}G_{F}N_{e}(r)}{\sqrt{(\Delta
    m^2)^2+(2E_{\nu}\sqrt{2}G_{F}N_{e}(r))^2-4\Delta
    m^2E_{\nu}\sqrt{2}G_{F}N_{e}(r)\cos2\theta}}.
\end{eqnarray} 
Assuming that $N_e\rightarrow0$ as $r\rightarrow\infty$, it is easy to
see that $|\nu_+(r\rightarrow\infty)\rangle\rightarrow|\nu_2\rangle$ and  
$|\nu_-(r\rightarrow\infty)\rangle\rightarrow|\nu_1\rangle$.
Therefore, if the transition from the core of the Sun to vacuum is
adiabatic, a state which is created as a $|\nu_+(0)\rangle$
($|\nu_-(0)\rangle$) will exit the Sun as a $|\nu_2\rangle$ ($|\nu_1\rangle$).

Having established the notation, it is very easy to estimate the
survival probability for electron neutrinos that are created in the
Sun's core and are detected on the Earth, in the limit that $\Delta
m^2/2E_\nu$ is much smaller than the Earth-Sun distance, such that
oscillations in vacuum between $\nu_1$ and $\nu_2$ states are
``averaged out.'' There are four possible 
``propagation paths'' that the solar neutrino can follow: 
\begin{eqnarray}
\nu_e &\rightarrow& \nu^+ (p=\sin^2\theta_M) \rightarrow \nu_2
(p=1-P_{c}) \rightarrow \nu_e (p=\sin^2\theta) \nonumber \\
 & \mbox{or} &  \nonumber \\
\nu_e &\rightarrow& \nu^+ (p=\sin^2\theta_M) \rightarrow \nu_1
(p=P_{c}) \rightarrow \nu_e (p=\cos^2\theta) \nonumber \\
 & \mbox{or} &  \label{paths} \\
\nu_e &\rightarrow& \nu^- (p=\cos^2\theta_M) \rightarrow \nu_1
(p=1-P_{c}) \rightarrow \nu_e (p=\cos^2\theta) \nonumber \\
 & \mbox{or} &  \nonumber \\
\nu_e &\rightarrow& \nu^- (p=\cos^2\theta_M) \rightarrow \nu_2
(p=P_{c}) \rightarrow \nu_e (p=\sin^2\theta), \nonumber
\end{eqnarray}
where $p$ is the probability that a given ``step'' takes place,
$\theta_M=\theta_M(0)$ and
$P_c$ is the jumping probability, {\it i.e.}\/ the probability that
during the evolution from the Sun's core to vacuum the neutrino
changes from one set of instantaneous Hamiltonian eigenstates to the
other.

Therefore, the probabilities of finding the mass eigenstates $\nu_1$ and
$\nu_2$ far from the Sun are given by
\begin{eqnarray}
  P_1 &=& \sin^2 \theta_M P_c + \cos^2 \theta_M (1-P_c) 
  \label{P1}\\
  P_2 &=& 1 - P_1,
\end{eqnarray}
where $\theta_M$ is that at the production point,\footnote{In our
  numerical analyses, we integrate over the production region using
  the profile given in \cite{bahcall_www}. The interference between
  $\nu_+$ and $\nu_-$ states in Eq.~(\ref{paths}) vanishes upon
  averaging over the production region independent of $\Delta m^2$ or
  energy. } 
and the electron neutrino survival probability ($P_{ee}$) at the
surface of the Earth is
\begin{eqnarray}
  P_{ee}&= &P_1 \cos^2 \theta + P_2 \sin^2 \theta \nonumber \\
&=& \sin^2\theta_M((1-P_c)\sin^2\theta+P_c\cos^2\theta)\nonumber \\
&+&\cos^2\theta_M((1-P_c)\cos^2\theta+P_c\sin^2\theta) .
\label{plus}
\end{eqnarray}
All equalities hold as long as two mass eigenstates appear as
an incoherent mixture (true for $\Delta m^2 \gtrsim 10^{-8}$~eV$^2$
for $^7$Be neutrinos).  
In deriving Eq.~(\ref{plus}), no assumption was made with respect to
the value of $\cos2\theta$, and therefore it should be valid for the
entire range of $0\leq\sin^2\theta\leq1$. 

Given Eq.~(\ref{plus}), it is easy to show that for
$\theta<\pi/4$, $P_{ee}$ can be (much) smaller than 1/2, while for
$\theta>\pi/4$, $P_{ee}$ is larger than 1/2 (indeed, it
will be shown that $P_{ee}\geq P_{ee}^{v}$, the (averaged) vacuum
survival probability).

First, note that $-1\leq\cos2\theta_M\leq\cos2\theta$. The equalities
are saturated when $\sqrt{2}G_{F}N_{e}(0)\gg\Delta m^2/2E_{\nu}$ or  
$\sqrt{2}G_{F}N_{e}(0)\ll\Delta m^2/2E_{\nu}$, respectively. More
quantitatively 
\begin{equation}
\frac{\Delta m^2}{2E_{\nu}\sqrt{2}G_FN_e(0)}= 0.98 \left(\frac{\Delta
    m^2}{10^{-5}\rm eV^2}\right)\left(\frac{0.862\rm MeV}{E_{\nu}}\right),
\end{equation}
for an average core electron number density of 79~moles/cm$^3$
\cite{SSM}. Therefore, in the case of $^7$Be neutrinos and 
$\Delta m^2\ll 10^{-5}$~eV$^2$, 
\begin{equation}
\cos2\theta_M=-1+ \frac{1}{2}\left(\frac{\Delta
 m^2\sin 2\theta}{2E_{\nu}\sqrt{2}G_FN_e(0)}\right)^2 + {\cal O}\left( 
 \frac{\Delta m^2}{2E_{\nu}\sqrt{2}G_FN_e(0)}\right)^3,
\end{equation}
and
\begin{equation}
P_{ee}\simeq (1-P_c)\sin^2\theta+P_c\cos^2\theta.
\end{equation}
We will soon show that $P_c\in[0,\cos^2\theta]$,\footnote{This is not
  hard to see. It is known that, if $\Delta m^2$ is large enough, the
  adiabatic approximation should hold, and therefore $P_c\rightarrow
  0$ for large enough $\Delta m^2$. On the other hand, if $\Delta m^2$
  is small enough, one should reproduce the vacuum oscillation result
  (as in the just-so scenario), and, from
  Eq.~(\ref{plus}), it is easy to see that this happens when
  $P_c\rightarrow \cos^2\theta$ and $\cos2\theta_M\rightarrow -1$.} so that, in the
limit $\cos 2\theta_M\rightarrow -1$, 
\begin{eqnarray}
\label{one}
P_{ee}\in [\sin^2\theta,\sin^4\theta+\cos^4\theta] & {\rm or} \\
\label{two}
P_{ee}\in [\sin^4\theta+\cos^4\theta,\sin^2\theta] 
\end{eqnarray}
Eq.~(\ref{one}) (Eq.~(\ref{two})) applies if $\sin^2\theta<\cos^2\theta$ 
($\sin^2\theta>\cos^2\theta$). This is easy to see because
$\sin^4\theta+\cos^4\theta=1-(1/2)\sin^22\theta$ is the vacuum
survival probability $P_{ee}^{v}$ and 
\begin{eqnarray}
P_{ee}^v&=&1-2\sin^2\theta(1-\sin^2\theta) \nonumber \\
&=& 1-2\sin^2\theta+2\sin^4\theta,
\end{eqnarray}
which is bigger (smaller) than $\sin^2\theta$ if $\sin^2\theta<\cos^2\theta$ 
($\sin^2\theta>\cos^2\theta$).

When $\sqrt{2}G_{F}N_{e}(0)\ll\Delta m^2/2E_{\nu}$ matter
interactions should be irrelevant, and it is easy to see from
Eq.~(\ref{theta_matter}) that $\cos2\theta_M\rightarrow
\cos2\theta$. In this limit $P_c\rightarrow 0$, since we are deep into
the adiabatic region (as will be shown later) and
$P_{ee}\rightarrow P_{ee}^v$. Before summarizing the behavior of the
$^7$Be electron neutrino survival probability we will determine some expression for $P_c$. 

Assuming an exponential profile for the electron number density inside
the Sun ($N_e(r)=N_e(0)\exp(-r/r_0)$), Schr\"odinger's equation can be
solved analytically \cite{Toshev, Petcov}, and it is shown that, 
in the range of the neutrino oscillation parameter space relevant for 
addressing the solar neutrino puzzle, Eq.~(\ref{plus}) is indeed a very
good approximation for $P_{ee}$ and that $P_c$ is given by 
\cite{Petcov,Petcov_Krastev} 
\begin{equation}
\label{Pc}
P_c=\frac{e^{-\gamma\sin^2\theta}-e^{-\gamma}}{1-e^{-\gamma}},
\end{equation}
where 
\begin{equation}
\gamma=2\pi r_0\frac{\Delta m^2}{2E_{\nu}}=1.22\left(\frac{\Delta m^2}
{10^{-9}\mbox{ eV}^2}\right)\left(\frac{0.862\mbox{ MeV}}{E_{\nu}}\right), 
\end{equation}
for $r_0=R_{\odot}/10.54=6.60\times 10^4$~km. 

According to the author of \cite{Petcov}, Eq.~(\ref{Pc}) only holds
for $\Delta m^2\cos2\theta>0$. 
We will prove shortly, however, that 
Eq.~(\ref{Pc}) also applies in the case of no level crossing, 
when the {\it heavy} mass eigenstate is predominantly of the electron 
type, {\it i.e.} when $\sin^2\theta>\cos^2\theta$. Assuming that this is
indeed the case, we can finish our discussion on the behavior of the
electron neutrino survival probability, using $^7$Be neutrinos as an
example.

When $\Delta m^2\ll10^{-9}$~eV$^2$, $\cos2\theta_M=-1$ and
$P_{c}=\cos^2\theta$. In this case we argued and one can explicitly
check that $P_{ee}=P_{ee}^v$.\footnote{Indeed, this is the region of
the ``just-so'' solution. As a matter of fact, in this region the
distance dependent vacuum oscillations do not average out when the
neutrinos are detected at the Earth, and one should use the
appropriate position dependent expression. It is known, however, that
the equality $P_{ee}(r)=P_{ee}^v(r)$ holds, up to a phase
\cite{seasonal,Pantaleone}. That this is also true for $\cos2\theta<0$
was explicitly checked starting with the exact solutions to
Schr\"odinger's equation \cite{Toshev,Petcov}.} For
$10^{-9}$~eV$^2\ll\Delta m^2\ll 10^{-5}$~eV$^2$, $\cos2\theta_M=-1$
and $P_{c}\rightarrow 0$. In this case $P_{ee}\simeq
\sin^2\theta$. This is the adiabatic region. For $\Delta
m^2\gg10^{-5}$~eV$^2$, matter effects become irrelevant and
$\cos2\theta_M=\cos2\theta$, $P_{c}=0$. Again
$P_{ee}=P_{ee}^v$. Therefore, Eqs.~(\ref{one},\ref{two}) apply for all
values of interest, and one can get a very large suppression of
$P_{ee}$ if $\sin^2\theta\ll 1$. On the other hand, in the case of no
level crossing, $P_{ee}$ is always bigger than $P_{ee}^{v}\geq 1/2$.

\begin{figure} [htbp]
\centerline{
  \psfig{file=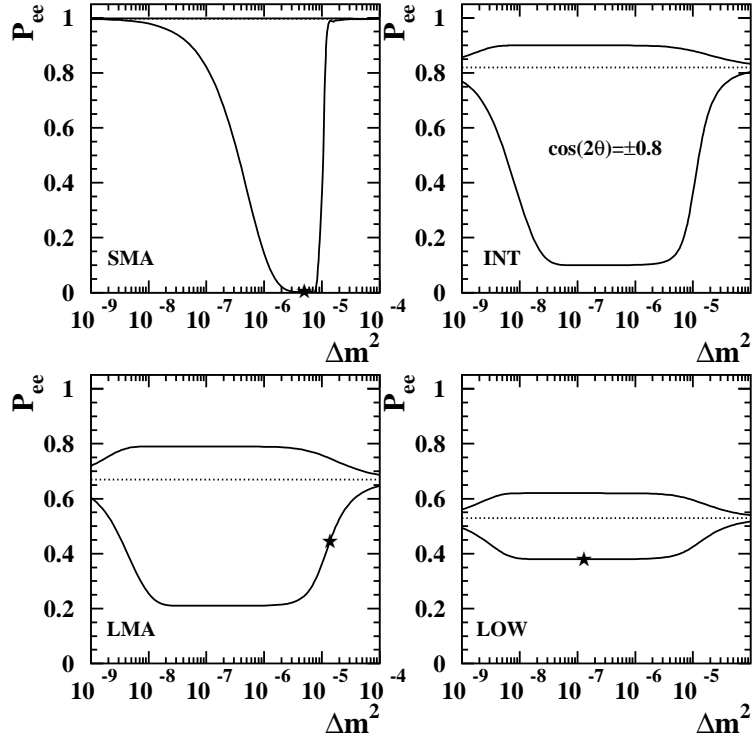,width=0.8\textwidth}
}
\caption{The electron neutrino survival probability as a function
  $\Delta m^2$, for different values of the vacuum mixing angle,
  namely, $\cos2\theta=\pm 0.997$ (SMA), $\cos2\theta=\pm 0.8$ (INT), 
  $\cos2\theta=\pm 0.58$ (LMA), and $\cos2\theta=\pm 0.24$ (LOW). The
  upper (lower) lines are for the negative (positive) sign of $\cos2\theta<0$. The
  stars indicate the preferred points from the overall rate analysis
  of the existing data \cite{rate_analysis}, and the horizontal dotted
  lines indicate the vacuum survival probability,
  $P_{ee}^v=1/2-1/2\sin^22\theta$ .}
\label{examples}
\end{figure}

\begin{figure} [htbp]
\centerline{
  \psfig{file=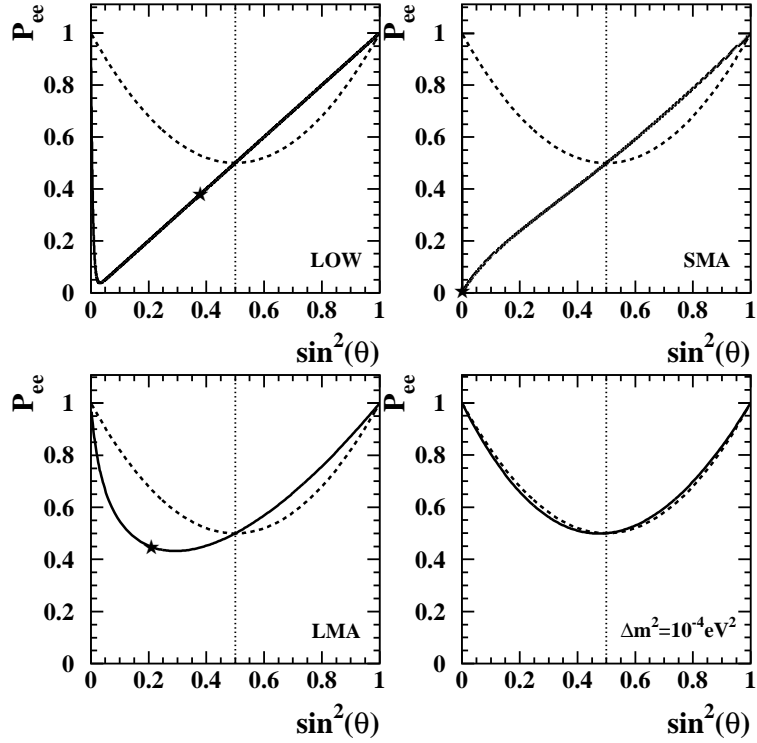,width=0.8\textwidth}
}
\caption{The electron neutrino survival probability as a function
  $\sin^2\theta$, for different values of $\Delta m^2$,
  namely, $\Delta m^2=1.3\times 10^{-7}$~eV$^2$ (LOW), $\Delta m^2=5.0\times
  10^{-6}$~eV$^2$ (SMA), $\Delta m^2=1.4\times 10^{-5}$~eV$^2$(LMA), and  
  $\Delta m^2=1\times 10^{-4}$~eV$^2$. The
  stars indicate the preferred points from the overall rate analysis
  of the existing data \cite{rate_analysis}, and the dashed
  lines indicate the vacuum survival probability,
  $P_{ee}^v=1/2-1/2\sin^22\theta$.}
\label{ex_angle}
\end{figure}

Fig.~\ref{examples} depicts the behavior of
$P_{ee}^{(^7\rm Be)}$ as a function of $\Delta m^2$, for different
values of the vacuum mixing angle. The preferred values from the
overall rate analysis at the Homestake, Sage and Gallex, and
SuperKamiokande experiments \cite{rate_analysis} are indicated by
stars. The four plots are labeled SMA, LMA, LOW to indicate that
they contain the best fit values of $\theta$ for the Small Mixing
Angle, Large Mixing Angle and LOW $\Delta m^2$ solutions
\cite{rate_analysis}, respectively, and INT to indicate an
intermediate value of $\theta$ between the SMA and LMA solutions. The
dotted line indicates the value of $P_{ee}^v$. 
Similarly, Fig.~\ref{ex_angle} depicts $P_{ee}^{(^7\rm Be)}$ as a
function of $\sin^2\theta$ for different values of the mass squared
difference. We use the same notation as the one used in
Fig.~\ref{examples}, and the vertical dashed lines indicate the mixing
angle for maximal vacuum mixing ($\sin^2\theta=1/2$). Note that at
this point $P_{ee}^{(^7\rm Be)}=P_{ee}^v=1/2$.  

Finally, we argue that Eq.~(\ref{Pc}) holds for all values of $\cos2\theta$.
When $\sqrt{2}G_{F}N_{e}(0)\gg|\Delta m^2|/2E_{\nu}$, it
is very simple to derive $P_{ee}$, following the exact solution
\cite{Toshev, Petcov} to
Schr\"odinger's equation and taking the appropriate limits. 
According to Eq.~(39) in \cite{Petcov}
\begin{eqnarray}
\label{Pmue}
P_{\mu e}&=&\frac{\sin^2(2\theta)}{4}\left[
\frac{\sinh(\pi r_0h_0\cos^2\theta)}{\cos^2\theta\sinh(\pi r_0h_0)}
e^{-\pi r_0h_0\sin^2\theta} \right. \nonumber \\
&+&\left.\frac{\sinh(\pi r_0h_0\sin^2\theta)}{\sin^2\theta\sinh(\pi r_0h_0)}
e^{\pi r_0h_0\cos^2\theta} + 
{\cal{O}}\left(\frac{\Delta m^2}{2E_{\nu}\sqrt{2}G_{F}N_{e}(0)}\right)^2 
\right], \\ 
P_{\mu e}&\simeq&
\sin^2\theta\left(\frac{\sinh(\pi r_0h_0\cos^2\theta)}{\sinh(\pi r_0h_0)}
e^{-\pi r_0h_0\sin^2\theta}\right) \nonumber \\
&+& \cos^2\theta\left(\frac{\sinh(\pi r_0h_0\sin^2\theta)}{\sinh(\pi r_0h_0)}
e^{\pi r_0h_0\cos^2\theta}\right), \nonumber \\
P_{\mu e}&\simeq&
\sin^2\theta\left(\frac{e^{\pi r_0h_0\cos2\theta}-e^{-\pi
      r_0h_0}}
{e^{\pi r_0h_0}-e^{-\pi r_0h_0}}\right) 
+\cos^2\theta\left(\frac{e^{\pi r_0h_0}-e^{\pi
      r_0h_0\cos2\theta}}
{e^{\pi r_0h_0}-e^{-\pi r_0h_0}}\right), \nonumber \\
P_{\mu e}&\simeq&
\sin^2\theta(P_c) 
+\cos^2\theta(1-P_c), \nonumber
\end{eqnarray}
where $h_0\equiv \Delta m^2/2E_{\nu}$ and $P_c$ is given exactly by
Eq.~(\ref{Pc}). Therefore
\begin{equation}
\label{fromexact}
P_{ee}=1-P_{\mu e}\simeq(1-P_c)\sin^2\theta+P_c\cos^2\theta.
\end{equation} 
Since, in deriving
Eq.~(\ref{Pmue}), no assumptions with respect to the sign of $\Delta
m^2$ or the value of $\theta$ were made, it should be applicable in
all cases,\footnote{That this is indeed the case was checked explicitly
  starting with the exact solution to Schr\"odinger's Equation in
  terms of Whittaker functions \cite{Toshev,Petcov}. Furthermore, in 
\cite{FLM}, the fact that Eq.~(\ref{Pc}) holds in the region of interested was 
verified numerically.} as long as 
$\sqrt{2}G_{F}N_{e}(0)\gg|\Delta m^2|/2E_{\nu}$.
Indeed, from Eqs.~(\ref{theta_matter}, \ref{plus}) it is easy
to note that, in the limit $\sqrt{2}G_{F}N_{e}(0)\gg|\Delta
m^2|/2E_{\nu}$, $\cos2\theta_M\rightarrow -1$ and Eq.~(\ref{fromexact})
is exactly reproduced. 

It should be noted that, in the region of no level crossing,
$P_c\approx 0$ for the entire $\Delta m^2$ range that we are interested
in. Indeed, only in the region of the ``just-so'' solution the
crossing probability is appreciably different from zero,  and it
  was in this region ($\cos2\theta_M=-1$) that we explicitly showed
  (Eq.~(\ref{Pmue}))  
  that Eq.~(\ref{Pc}) holds. If there are any
interesting phenomenological consequences for the
$\theta>\pi/4$ case 
in the ``just-so'' to MSW transition region of the parameter space
still remains to be seen \cite{new}.

\end{document}